# Investigation on the Mechanical Properties of Functionally Graded Nickel and Aluminium Alloy by Molecular Dynamics Study

Shailee Mitra[1], Md. Habibur Rahman[1], Mohammad Motalab[1*], Tawfiqur Rakib[2], Pritom Bose[1]

[1]Department of Mechanical Engineering, Bangladesh University of Engineering and Technology, Dhaka-1000, Bangladesh

[2]Department of Mechanical Science and Engineering, University of Illinois at Urbana-Champaign, Urbana, IL 61801, United States

*Corresponding Author. E-mail address: mtipuz@yahoo.com

## Abstract

Functionally graded materials (FGMs), have drawn considerable attention of the worldwide researchers and scientific community because of its unique mechanical, thermal and electrical properties which may be exploited by varying the compositions gradually over volume. This makes FGM multifunctional material (properties changing continuously in a certain direction) for specific purpose without creating any phase interface thus making it superior to its composite counterparts. In this paper, we applied Molecular Dynamics (MD) approach to investigate the mechanical properties of functional graded Ni-Al alloy with Ni coating by applying uniaxial tension. Nickel-Aluminum (Ni-Al) alloy has been used extensively in the industry due to its remarkable mechanical and thermal properties. Our aim is to find the difference in material behavior when we change the grading function (linear, elliptical and parabolic), temperature and crystallographic direction. We also observe distinct type of failure mechanism for different grading function at different temperature. Close observation reveals that elliptically graded Ni-Al alloy has high tensile strength at low temperature whereas at high temperature, the highest tensile strength is found for parabolic grading. Besides, at any temperature, the parabolically graded Ni-Al alloy shows superior elasticity than its elliptical and linear counterpart. Moreover, it is also observed that [111] crystallographic direction for this alloy demonstrates more resistivity towards failure than any other crystallographic direction. It is found that lattice disorder plays a significant role on the mechanical properties of Functionally Graded Materials (FGMs). This paper details a pathway to tune the mechanical properties like Young's Modulus, plasticity and yield strength at molecular level by varying the composition of materials along different grading functions.

## 1. Introduction

Functionally Graded Materials (FGMs) have great potential for applications in as heat-resistant materials but also as structural materials, biomaterials, semiconductors, and electrode materials [1-3]. The concept of FGMs was first initiated in Japan to use in high temperature applications like rocket engine, jet engine and others aerospace application which can carry high temperature gradient and to reduce thermal stress [4]. According to Udupa et al. functionally graded materials are inhomogeneous materials consisting of two or more different materials, engineered to have continuously varying spatial compositions for specific application [1]. Tarlochan reported that FGM are the advanced materials in the family of engineering composites made of two or more constituent phases with continuous and smoothly varying compositions [2]. According to him, FGMs are engineered based on different gradients of composition in the preferred material axis orientation, thus making it flexible, and superior to homogeneous material composed of similar constituents [2]. FGM coating can reduce the crack driving force to connect the materials to eliminate the stress at interface and can greatly enhance bond strength [6]. Different techniques have been used to manufacture bulk FGM like CVD (Chemical Vapor Deposition) process, PVD (Physical Vapor Deposition) process, surface reaction process and last of all plasma spraying process [7-12]. Different experimental and numerical approach have been conducted to investigate the characteristics of FGM at bulk and micro level [13-18]. Cui et al. investigated on optimizing the design of functionally graded material layer in all-ceramic dental restorations [14]. They reported that in natural tooth structure, stress concentration is reduced by the functionally graded structure of dentin-enamel junction which interconnects enamel and dentin and this concept can be applied in all-ceramic dental restorations to achieve excellent stress reduction and distribution [14]. However, FGM at nanoscale has not been explored extensively yet. In this research work we aim to investigate the temperature dependent mechanical properties of functionally graded Ni and Al alloy. For our study, we choose tri-nickel aluminide ($Ni_3Al$, IC-221M super alloy) which is an intermetallic alloy of Ni and Al having properties similar to both a ceramic and a metal, thus making it ideal for FGM [1-3,5]. $Ni_3Al$ is unique in that it has very high thermal conductivity combined with great mechanical strength at high temperature [19-26]. According to Jozwik et al. $Ni_3Al$ has high tensile and compression strength at temperature of 650-1100 °C, high corrosion resistance in oxygen and carbon enriched atmosphere, high creep and fatigue strength, remarkable wear resistance at high temperature and a relatively low density which gives it high strength to weight ratio [19]. These properties and its low density make it ideal for special applications where superior strength is required at higher temperature and corrosive environment [27-31]. Molecular dynamics approach is proven to be effective to find out the temperature dependent mechanical properties of Al and Ni at nanoscales their different intermetallic alloy [32-40]. But to the best our knowledge, the mechanical properties of $Ni_3Al$ at atomic level has not been reported yet. Yang et al. investigated melting characteristics of $Ni_3Al$ by molecular dynamics simulation using EAM potential [40]. They reported high melting point of $Ni_3Al$ alloy [40]. This paper presents atomistic simulation results of tensile tests on Ni-Al FGM alloy based on different profiles such as linear, elliptical and parabolic at various temperature. We are interested in Ni-Al FGM because material properties especially heat, corrosion resistance and mechanical characteristics can be tuned along target specific dimensions. The variance of compositions are varied in particular direction which makes FGM more interesting material compared to their conventional alloy as reported earlier [1-3,5]. Therefore, the gap in the understanding of the effect of profile, temperature and crystal orientations on the mechanical properties of Ni-Al FGM necessitates a comprehensive study. The FGM nanostructures of Ni-Al considered in this investigation are subjected to uniaxial tensile strain, where the temperature is varied from 100K to 600K and three crystal orientations [100], [110] and [111] have been chosen. The stress strain relationships are reported for tensile loading with variation with profile, temperature and crystallographic orientation. The variations of ultimate tensile strength, Young's modulus based on various profile with temperature and crystallographic orientation have been discussed in-depth. Moreover, the temperature dependent failure mechanism of linearly graded Ni-Al has been elucidated in this paper.

## 2. Computational Method

The lattice of perfect single crystal Al and Ni is FCC (face centered cubic) structure with a lattice parameter of 4.06 $\dot{A}$ and 3.52 $\dot{A}$ respectively [33,35]. In Fig.1, we showed lattice structure for the single crystal $Ni_3Al$ is $L_{12}$ having a lattice parameter of 3.57 $\dot{A}$ [21] where Ni atoms are on face centers and Al atoms at corners [21]. Variation in percentage of Al atom in Ni matrix is introduced by replacing Al with Ni from the corner of the $Ni_3Al$ crystal by following different grading function linear, elliptical and parabolic with the help atomsk [51].

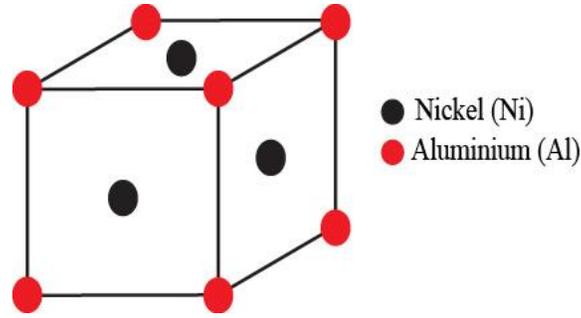

**FIGURE 1.** Ni$_3$Al (L$_{12}$) crystal structure

The dimension of the Ni-Al FGM structure is taken as 28.57 nm x 3.57 nm x 3.57 nm which maintains a length to width ratio of 8:1 [45,46]. Three different crystallographic orientations ([100], [110] and [111]) are studied. In Fig. 2, we showed different types of grading function employed in the Ni-Al FGM nanostructure. In Fig. 3, the initial structure of FGM alloy with linear, parabolic and elliptic grading have been shown.

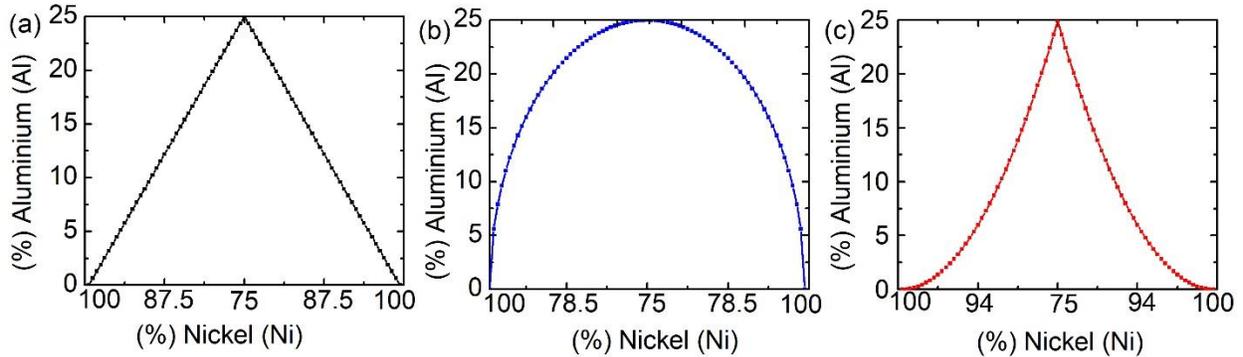

**FIGURE 2**. (a) Linear grading (b) Elliptical grading and (c) Parabolic grading function

Molecular dynamics calculations of mechanical properties are carried out using Large-scale Atomic Massively Parallel Simulator (LAMMPS) [49] and visualization of atomistic deformation processes is done by OVITO [50] .This work utilizes Embedded Atom Method (EAM) potential [36] to describe the interactions between the Al-Al, Ni-Ni and Ni-Al atoms in molecular dynamics simulation. This EAM potential is widely used to define atomic interactions for Aluminum, Copper, Nickel, Iron and other metals [36,42,43]. In our simulation, time step is chosen as 1 fs [45,46]. At first, energy of the system is minimized using Conjugate Gradient (CG) minimization scheme [48]. Before applying the tensile load, NVE and NPT equilibration are carried out for 100 ps and 100 ps respectively [45,46]. Finally, uniaxial tensile strain is applied along X direction at a constant strain rate of $10^9 \, s^{-1}$ under NVT ensembles to control temperature fluctuations [46]. Periodic boundary conditions are applied at loading direction (X) and other directions are kept free so that we can get the effect of surface of the nanostructure [45]. This strain rate is higher than that adopted in real life because we are constrained by the computational resources. But previous works have found that the strain rate of $10^9 \, s^{-1}$ is good enough for atomistic simulations [44-47]. The atomic stress in our simulation is calculated using Virial stress theorem [46] which stands as,

$$\sigma_{virial}(r) = \frac{1}{\Omega}\sum_i(-\dot{m}_i\dot{u}_i \otimes \dot{u}_i + \frac{1}{2}\sum_{j\neq i} r_{ij} \otimes f_{ij})$$

In this theorem, the summation is performed over all the atoms occupying the total volume, $\dot{m}_i$ presents the mass of atom $i$, $\otimes$ indicates cross product, $\dot{u}_i$ is the time derivative, which indicates the displacement of the atom with respect to a reference position, $r_{ij}$ represents the position vector of the atom, and $f_{ij}$ is the interatomic force applied on atom $i$ by atom $j$. [46].

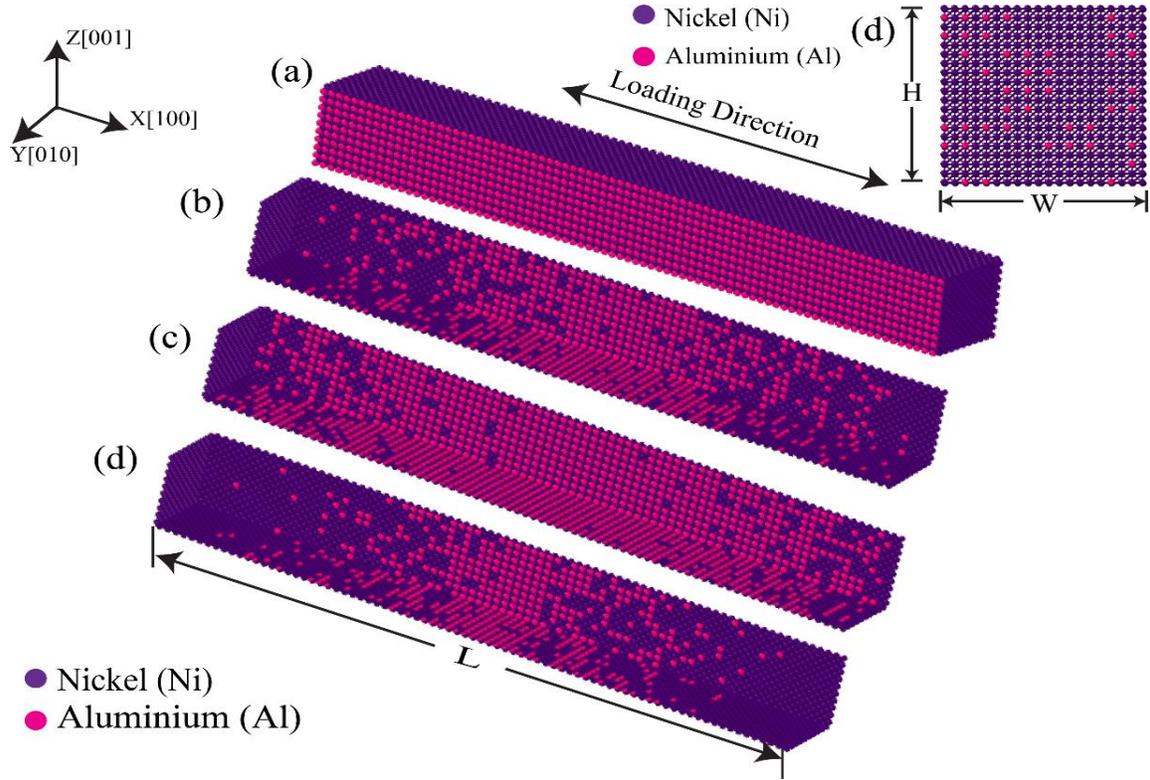

**FIGURE 3**. Initial coordinates of Ni-Al nanostructure of (a) Ni$_3$Al alloy, FGMs with (b) Linear grading (b) Elliptical grading (c) Parabolic grading and (d) Cross section of linearly graded Ni-Al alloy.

To check the validity of our computational code and potential used, Young modulus of [100] oriented Ni, Al and melting point of [100] oriented Ni$_3$Al is compared with published literatures [33,36,40] which are shown in Fig.4 (a-b). To determine the melting point of Ni$_3$Al, equilibration was performed in the NVT ensemble in two steps of 50 ps each, constraining the position of the center of mass of the atoms by adjusting the coordinates. The system was then placed into the NPT ensemble, where the temperature of the system is increased at 1 bar from 300K to 2500K [41]. Additionally, the change in the total energy of Ni$_3$Al during the temperature rise was analyzed to get a clear idea on the melting point [41]. Fig. 4 (b) demonstrates the melting point by showing the change in the total energy of the Ni$_3$Al with temperature. The bend in the energy curve corresponds to a first order melting transition of the material [41]. Young's Modulus is calculated from the stress-strain curve and strain value less than 4% using linear regression [45.46]. We could not compare any mechanical properties with the existing literature for Ni$_3$Al as the mechanical properties of this alloy has not been studied extensively. However, we compared the melting point of the system which establishes the validity of the use of our potential to describe the interactions between the atoms in a dynamic system.

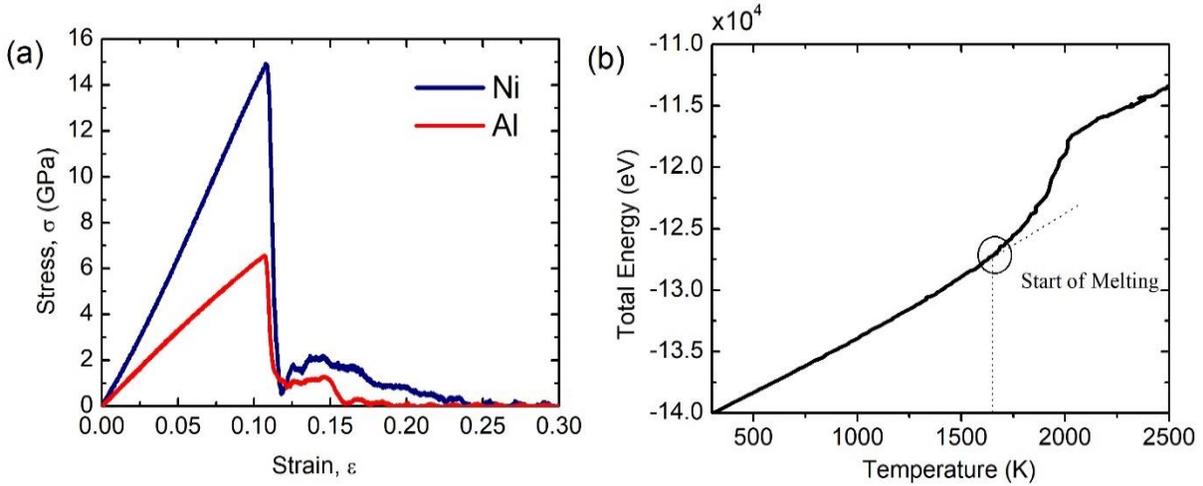

**FIGURE 4**. (a) Stress-strain curve of Ni and Al for 28.57 nm x 3.57 nm x 3.57 nm at 300K  (b) Determination of the melting point of $Ni_3Al$ .

**TABLE 1**: Comparison between values obtained by the present method and existing literature

| Material | Young's Modulus (GPa) | Melting Point (K) |
|---|---|---|
| Al | 68 (GPa) [Present calculation] 69 (GPa) [33] | ----- |
| Ni | 130 (GPa) [Present calculation] 116.5 -119.4 (GPa) [36] | ----- |
| $Ni_3Al$ | ------ | 1655K [Present calculation] 1663K [40] |

So, we can conclude that the potential used to define the interatomic interaction in this study can accurately represent the mechanical properties of the Ni-Al FGM alloy.

# 3. Results and Discussions

**3.1 Impact of temperature, profile and comparison between $Ni_3Al$ alloy with FGMs:** Fig.5 (a-c) represent stress-strain response of $Ni_3Al$, FGM Linear, FGM Elliptical and FGM Parabolic at 100K, 300K and 600K respectively. The main target of our study was to determine stress-strain behavior of various FGMs profile at different temperature. MD (Molecular Dynamics) simulation has been performed at 100K to 600K. Fig. 5(d-e) show the variation of ultimate tensile stress and Young's Modulus respectively as a function of temperature for different grading function. From the stress-strain diagram in Fig. 5(a-c), it is clear that the materials ($Ni_3Al$, FGM Linear, FGM Elliptical and FGM Parabolic) exhibit ductile type failure. At the early stage of tensile deformation (strain less than 4%) the stress strain relation is almost linear and the structures (FGM Linear, FGM Elliptical, FGM Parabolic) exhibits elastic behavior.

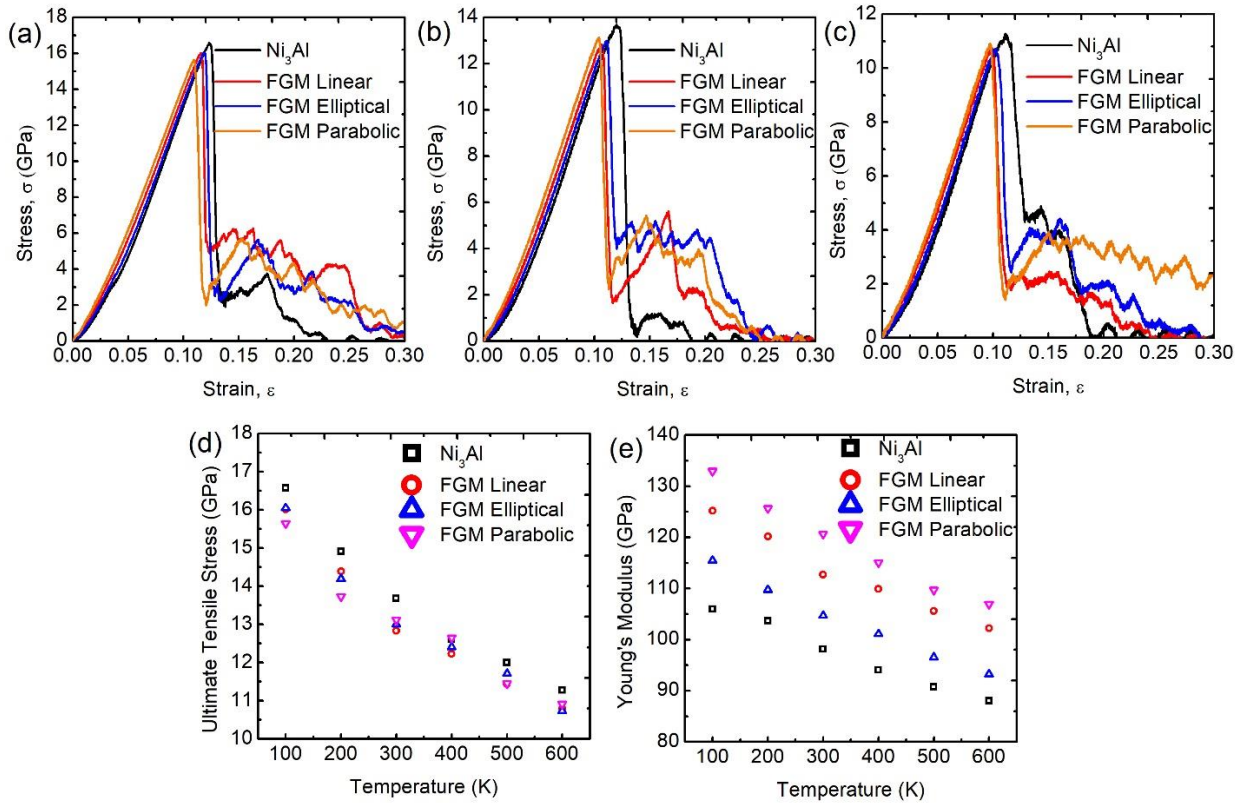

**FIGURE 5**. (a) Stress-strain curve of different FGM nanostructure at (a) 100K (b) 300K (c) 600K. (d) Variations of ultimate tensile stress for different FGM nanostructure with temperature and (e) variations of Young's Modulus for different FGM nanostructure with temperature.

This elastic region is approximated by a linear regression (strain less than 4%) and then the Young's Modulus is determined by applying a curve fit to the corresponding generated data [43,46]. Then the stress increases nonlinearly until it reaches a peak value known as ultimate tensile stress and then the magnitude of stress suddenly decreases. Due to ductile nature of Ni-Al as reported earlier [32-39] and its alloy [39] the materials do not fail catastrophically. That's why, In the stress-strain graphs, flow stress is observed which originates from the plasticity in the metallic structures [42,43]. The zig-zag portion of the curve, known as plastic stress flow, is due to nucleation, transformation and junction of dislocations in materials [42,43]. Due to the tensile load, dislocations formed at different regions interact and annihilate each other which leads to the final failure of the material [42,43]. It can be easily be visualized from Fig. 5(d) and Fig. 5(e) that at higher temperature materials experience higher thermal fluctuation and lattice vibration which causes materials to fail at a low strain. That's why, the ultimate tensile strength and the Young's Modulus show a decreasing trend with increasing temperature. This kind of trend is previously seen in other materials too [42-46]. From Fig. 5 (d), we can see that $Ni_3Al$ is showing high ultimate tensile strength compared to FGM profiles. That is because there is no lattice mismatch in $Ni_3Al$. We know that $Ni_3Al$ has a lattice constant of 3.57 $\mathring{A}$ with an FCC ($L_{12}$) crystal structure as mentioned earlier. The grading is introduced in this material by replacing Al atoms from the corner by Ni atom. When we design the Ni-Al FGM nanostructure in this way, this 3:1 ratio of Ni and Al atoms is not maintained. The grading introduces inhomogeneity in the ratio of Ni and Al in the different plane of the alloy. For this reason, a lattice mismatch [42] is introduced in the nanostructure. For elliptical, linear, and parabolic profiles, the lattice mismatch is calculated as 1.13%, 1, 24%, and 1.30% respectively. At low temperature, say 100 K, the effect of lattice mismatch dominates the stress-strain relation. That's why, the ultimate stress for parabolic grading is the least and that of elliptical grading is the highest. However, Closer inspection reveals that ultimate tensile strength of elliptical profiles is the least at very high temperature. That is because while Ni atoms are always hard, Al atoms become soft at high temperature. In case of elliptical grading, the percentage of Al atoms is higher than the other profiles. At high temperature, the softness of Al atoms dominates over the lattice mismatch and therefore, the ultimate stress of Ni-Al alloy with elliptical grading becomes the least. On the other hand, In Fig. 5(e), we can see that the Young's Modulus is higher for parabolic grading than the other profiles. Ni-Al FGM alloy system with parabolic grading has higher Ni percentage than the others. As Ni is more elastic than Al [33,36], parabolic profile shows more elasticity. Moreover, Young's modulus is calculated as a linear elastic property in this study and the effect of lattice mismatch is non-linear. That's why, there is no effect of lattice mismatch in Young's modulus. Since the ultimate stress is found in the non-linear region of stress-strain, the effect of lattice mismatch dominates more in the trend of ultimate tensile stress rather than Young's modulus. It is also evident from Fig. 5(a-e) that FGM profiles showed better Young's Modulus, failure resistance and ductility compared to $Ni_3Al$ alloy which is in good agreement with previous experimental research [1-4] regarding superiority of FGMs over conventional homogenous alloys. From our results, we calculated flow stress as 0.35, 2.25, 3.5 and 2.8 GPa for $Ni_3Al$, linear, elliptical and parabolic FGM respectively at 300K thus indicating better ductility in FGM profiles compared to $Ni_3Al$ alloy. Flow stress is calculated from the average values of stress corresponding to strain 0.14-0.25 [42,43].Thus, we can also conclude that grading the Ni-Al alloy functionally can be an effective way to enhance mechanical properties of the alloy.

**3.2 Impact of crystal direction on linearly graded FGM:** Figure 7(a) shows the stress-strain curves of Ni-Al FGM alloy with linear grading at 300K temperature with [100], [110] and [111] crystal orientations. It can be observed that the ultimate tensile stress of the [111] orientation is the highest and for [110] orientation, it is the lowest. The flow stress calculation in different crystal orientation for linearly graded FGM shows that [110] and [111] crystal orientation has better ductility than [100] crystal orientation. Fig. 7 (b) shows variations of ultimate tensile stress with temperature. It has been clearly seen that [111] crystal direction shows the largest ultimate tensile stress than the other crystallographic directions. It has been reported earlier that [111] crystal orientation has the lowest surface energy [48,52] so it has the highest tensile strength and then comes [100] and [110] crystal orientation respectively for fcc lattice structure which is in agreement with our simulations results [48]. In case of [111] direction the atoms are arranged is such a way which has been shown in Fig.6 (f), it requires higher energy to break the bond between Ni and Al as bond direction makes an angle of $45^0$ with loading direction. From Fig.6 (b) and Fig.6 (d) it can be seen that metallic bond length between Al and Ni in loading direction is larger in case of [110] orientation. So it takes less energy to break the bond of [110] crystal orientation compared to [100] orientation. Therefore, the ultimate tensile stress for [110] direction is low compared to that of [100] orientation. Closer study reveals that rate of decrement of ultimate tensile stress with temperature is lower in case of [110] crystal orientation. When temperature increases from 100K to 600K, ultimate tensile stress of [110] crystal orientation is reduced by 28%. But rate of decrement of ultimate tensile stress is larger for [111] crystal orientation. When temperature increases from 100K to 600K ultimate tensile stress of [111] crystal orientation reduced by 35%. It may be attributed from the non-equilibrium thermodynamic states of [111] crystal orientation at high temperature. At high temperature, interatomic bond strength between Ni-Al along [111] orientation weakens resulting in a greater decrement of ultimate tensile strength for [111] crystal orientation. Fig. 7 (c) shows variations of Young's Modulus with temperature. Young's Modulus is calculated from stress-strain curve and strain value less than 4% as mentioned earlier. Form Fig. 7(c) it can be easily seen that [111] crystal direction has the highest Young's modulus making it more elastic than the other crystallographic directions [52]. Though [110] orientation has lower ultimate tensile stress compared to [100] orientation. In case of Young's Modulus, [110] orientation shows more elasticity than [100] crystal orientation. It may be attributed from sharp rise in stress at lower strain value for [110] crystal orientation compared to [100] crystal orientation. In [110] direction, due to the atomic arrangements in the crystal plane, there is a fast accumulation of stress when the tensile load is applied. That's why, the fast accumulation leads to higher Young's modulus than [100] direction. Rate of decrement of Young's Modulus for [111] orientation is larger compared to other crystal directions. When temperature increased from 100K to 600K Young's modulus of [111] crystal orientation reduced by 18.30 %. So it can be concluded that target specific crystal growth can be a way to tune the mechanical properties while manufacturing FGM at both micro and bulk level.

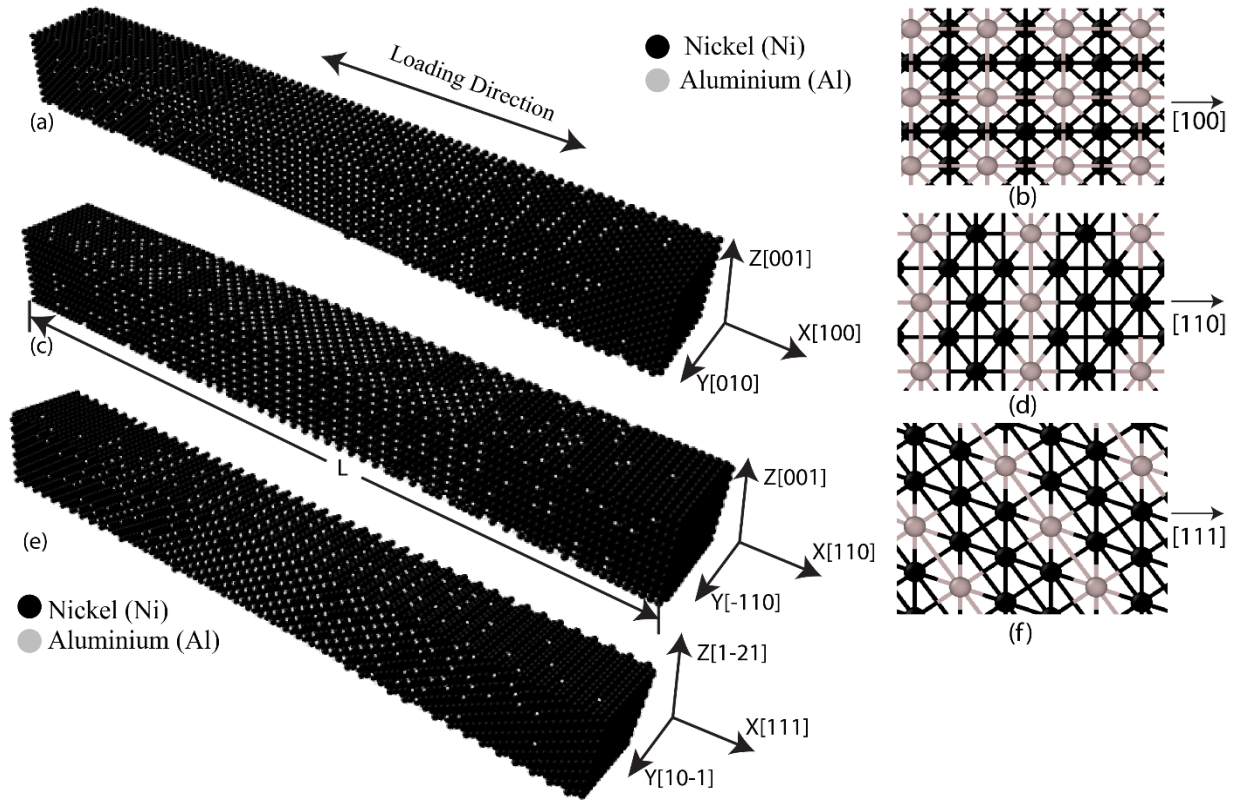

**FIGURE 6**. Initial nanostructure of FGM Linear (a) [100] (c) [110] (e) [111] crystal direction. Atomistic representation of FGM linear along (b) [100] (d) [110] [f] [111] crystal direction.

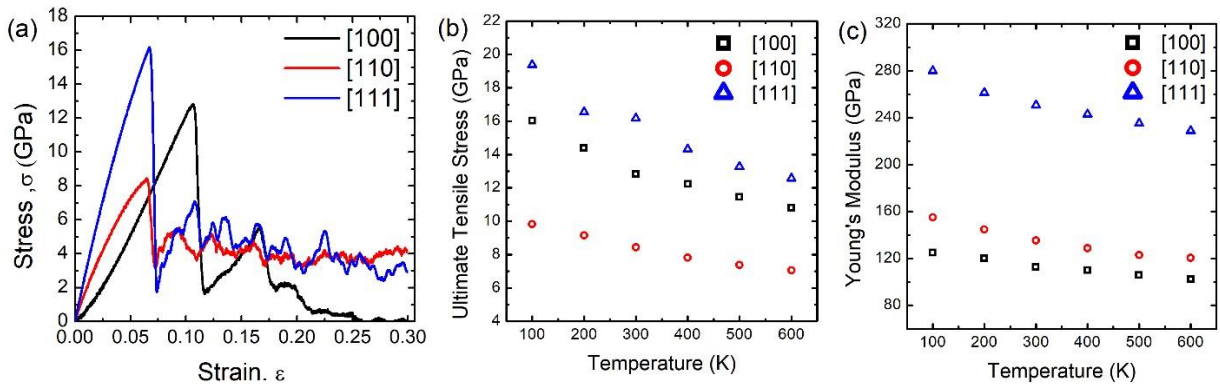

**FIGURE 7**. (a) Stress-strain curve of FGM linear for different crystal orientation at fixed 300K. (b) Variations of ultimate tensile stress of FGM linear for different crystal orientation with temperature and (c) variations of Young's Modulus of FGM linear for different crystal orientation with temperature.

**3.3 Failure mechanism of [100] oriented linearly graded FGM at 100K and 400K:** Fig. 8(a) and 8(b) show failure mechanism of [100] crystal orientated FGM Linear at 100K and 400K respectively for various strain level in terms of shear strain parameter.

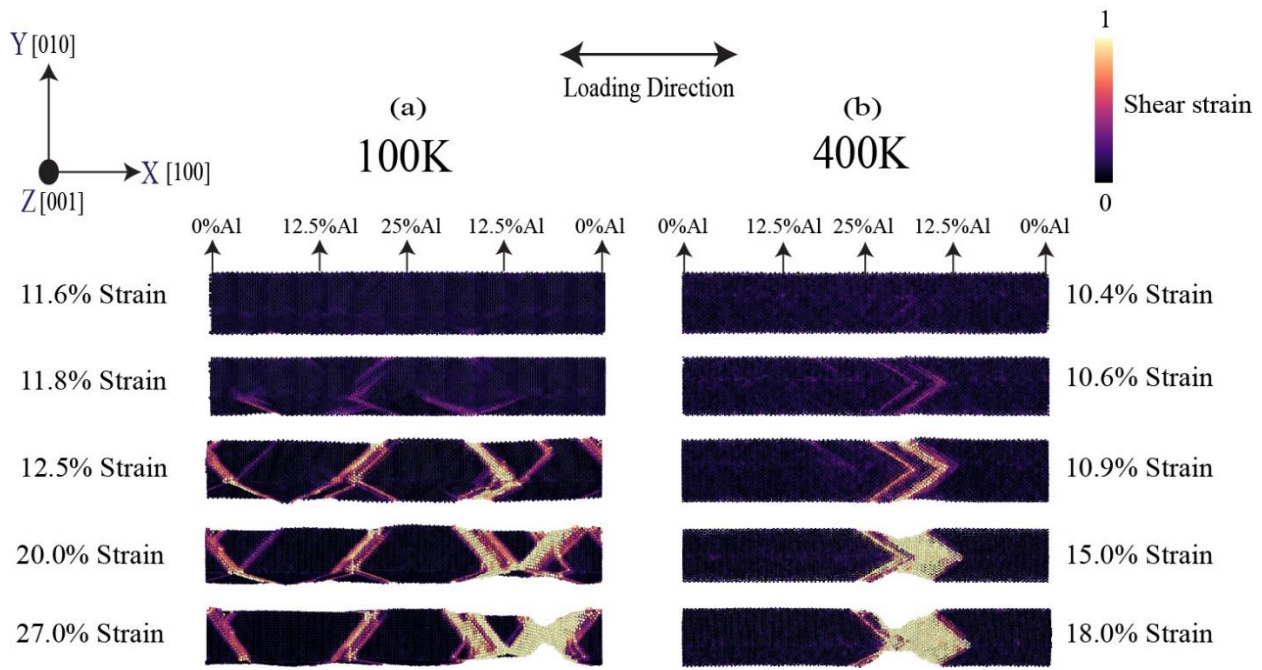

**FIGURE 8**. (a) Failure mechanism of [100] oriented FGM Linear (a) 100K and (b) 400K under uniaxial tensile simulation for different strain level.

Shear bands are formed at 11.8% strain and 10.6% strain for 100K and 400K respectively. Interestingly, it is also observed that at low temperature, the shear bands form in a distributed way whereas at high temperature, the shear bands are concentrated in a definite position. This causes an increase in the local stress concentration. This leads to early failure at high temperature. However, in the case of low temperature, the distributed shear bands doesn't allow stress build-up in a local region at a low strain. Because of the low temperature, the atom's fluctuation from its mean position is not very rapid. However, as the strain keeps increasing, the atom moves from its equilibrium and the local pile-up of shear bands is evident which ultimately causes failure. It can also be visualized from Fig. 8(a) that at 27% strain, stress concentration causes necking. The necking takes place in a region where percentage of Al is very low, about 6.5% and lattice disorder in this region is high. On the other hand, at 400 K, the stress concentration and necking occurs near the middle region where the percentage of Al is about 18.5%. We already know that Al becomes softer at high temperature. That's why, the failure occurs by necking in Al enriched region of the alloy. So it can be concluded that at low temperature failure behavior is mainly dominated by lattice mismatch effect but at high temperature failure mechanism is dominated by interatomic bond strength between Ni-Al, Ni-Ni and Al-Al.

## 4. Conclusions

We have studied the effect of profiles, temperature, crystal orientations on the mechanical behavior of functionally graded Ni-Al alloy with Ni coating. On the basis of our Molecular Dynamics (MD) simulations it is found that the stress –strain curves show ductile failures for $Ni_3Al$, linearly, parabolically, and elliptically graded Ni-Al alloy. As temperature increases from 100K to 600K, mechanical properties such as ultimate tensile strength, and Young's Modulus and failure strain decreases thus showing an inverse relationship with temperature. The main findings of this study are FGM profiles show better failure resistance, Young's Modulus and ductility compared to $Ni_3Al$ alloy. This proves that functionally grading an alloy can be an effective way to enhance materials properties. It is also found that [111] crystallographic orientated linearly graded FGM shows better ultimate tensile strength and elasticity compared to other crystallographic orientations. We have concluded that at low temperature lattice mismatch dominates the failure mechanism of FGM profiles, but at high temperature due to softness of Al, the effect of lattice mismatch is low. Materials fails differently on the basis temperature which have been elucidated in this article. Overall, this article can serve as a comprehensive way to learn the details of failure of Ni-Al FGM alloys and how we can tune the mechanical properties of Ni-Al alloy by functional grading.

## Acknowledgement

Authors of this article would like to thank Department of Mechanical Engineering, Bangladesh University of Engineering and Technology (BUET) and Multiscale Mechanical Modeling and Research Networks (MMMRN) for providing computational resources.